\documentclass[10pt,showpacs,letterpaper,aps,pra,twocolumn]{revtex4-1}
\usepackage[draft]{hyperref}
\usepackage{braket}
\usepackage{amsmath}
\usepackage{graphicx}
\usepackage{amssymb}
\usepackage{esint}
\usepackage{comment}

\newcommand{\uu}{\mathcal{U}}

\newcommand{\hc}{\text{h.c.}}
\newcommand{\sinc}{\text{sinc}}
\newcommand{\cc}{\text{c.c.}}
\newcommand{\vac}{\text{vac}}

\begin{document}
\preprint{APS/123-QED}

\title{Effects of time ordering in quantum nonlinear optics}
\author{Nicol\'as Quesada and J. E. Sipe}
\affiliation{McLennan Physical Laboratories, University of Toronto, 60 St. George Street, Toronto, ON, M5S1A7, Canada }
\email{nquesada@physics.utoronto.ca}

\begin{abstract}
We study time ordering corrections to the description of spontaneous parametric down-conversion (SPDC), four wave mixing (SFWM) and frequency conversion (FC) using the Magnus expansion. Analytic approximations to the evolution operator that are unitary are obtained. They are Gaussian preserving, and allow us to understand order-by-order the effects of time ordering.  We show that the corrections due to time ordering vanish exactly if the phase matching function is sufficiently broad. The calculation of the effects of time ordering on the joint spectral amplitude of the photons generated in spontaneous SPDC and SFWM are reduced to quadrature.
\end{abstract}

\maketitle

\section{Introduction}
\noindent Spontaneous parametric down-conversion (SPDC), spontaneous four-wave mixing (SFWM), and frequency conversion (FC) are three of the most common nonlinear
processes used in quantum optics.  In the first two a nonlinear medium is used to convert photons from a pump laser into pairs of quantum correlated photons \cite{christ13}.  In the third, the frequency of a photon is increased or decreased after it interacts with a strong pump field inside a nonlinear medium.  Pair generation is often modeled in a straightforward manner using the first order of a Dyson series, or the first term in a Taylor-like series that results if consequences of time-ordering are ignored \cite{grice97}.  But when the nonlinear interaction becomes sufficiently strong a more involved theoretical treatment is required. In particular, the Taylor-like series is suspect beyond a few orders because  the interaction Hamiltonian does not commute with itself at different times.

In previous work \cite{aggie10,aggie11} the Dyson series has been used to investigate parametric down-conversion, and it was shown that when the initial state of the down-converted modes is vacuum (SPDC), the Taylor-like and Dyson series give identical results up to second order. In a subsequent study, Christ et al. \cite{christ13n} used the Heisenberg picture to investigate time ordering issues in both SPDC and FC. They noticed that when the pump is approximated as undepleted and treated classically, the Hamiltonian is quadratic in the bosonic operators of the down-converted fields, and thus the equations of motion of the operators are necessarily \textit{linear.}  This implies that the outgoing fields are related to the incoming fields by a linear Bogoliubov transformation \cite{Braunstein05}.  They used this observation to build an \textit{ansatz} solution, and develop a numerical algorithm that permits the calculation of the linear transformation relating output to input.  Another consequence of the linear Bogoliubov relation between output and input is that the state of the down-converted photons in SPDC cannot be more complicated than a two-mode squeezed vacuum. This is at odds with a result of Branczyk et al. \cite{aggie11}, who found that the third order correction to the SPDC state contains a six photon energy correlated state. As we shall show, this discrepancy is
related to the fact that the Dyson series does not retain the Gaussian preserving nature of the quadratic Hamiltonian, or equivalently the linear Bogoliubov nature of the transformation relating input and output fields.

In this paper we employ the Magnus Expansion (ME) perturbation series, which is able to deal properly with time ordering at each level, and thus is unitary to all orders. \ Yet, unlike the Dyson series, it respects the Gaussian preserving nature of the process for the class of Hamiltonians that are quadratic in bosonic operators, giving rise to a linear Bogoliubov transformation to whatever order it is taken.  Thus we feel it should be the preferred strategy for the description of photon generation and conversion in nonlinear quantum optics.

In Section II we introduce the ME and mention some of its important properties, deriving a useful simplification of its terms when dealing with a broad class of Hamiltonians used in quantum optics; we also show why the ME automatically respects the Gaussian nature of usual nonlinear quantum optical processes.  In Section \ref{sec:pdc} we apply the Magnus series generated by the ME to SPDC and SFWM Hamiltonians.  While the ME has been used in the past in quantum nonlinear optics (see e.g. Yang et al. \cite{Yang08} and references cited therein), usually only the first term in the expansion has been kept. We calculate the second and third order terms, and also introduce diagrams akin to Feynman diagrams that allow for a simple tabulation of the higher order processes in the ME. Using these diagrams we  easily demonstrate why it is necessary to go to third order in the perturbation expansion to see a correction to the SPDC and SFWM states. In section \ref{sec:fc} we extend these ideas to FC processes. We show that the expansion formulas derived for FC are completely analogous to those derived for SPDC and SFWM. In section \ref{sec:dis} we show how the ME can be taken to a form more amenable to direct calculation and develop a perturbation series for the joint spectral amplitude (JSA) of the down converted photons in SPDC and SFWM. In section \ref{sec:broad} we show that for perfect phase matching the second and third order Magnus terms vanish. Finally, conclusions and remarks are presented in section \ref{sec:end}.

\section{Properties of the Magnus expansion}\label{sec:magnus}
\noindent In quantum mechanics it is often necessary to solve the time dependent Schr\"{o}dinger equation for the time evolution operator $\uu(t,t_0)$,
\begin{eqnarray}\label{schro}
i\hbar \frac{d}{dt} \uu (t,t_0)=H_I(t) \uu (t,t_0),
\end{eqnarray}
where $H_I(t)$ is the interaction picture Hamiltonian.
One common way to approximate the solution of Eq. (\ref{schro}) is to use the Dyson series in which the evolution operator is expanded as a power series
\begin{eqnarray}\label{dysondef}
\uu(t_1,t_0)&=& \mathbb{I}+T_1+T_2+\ldots\\
&=&\mathbb{I}+(-i)\int_{t_0}^{t_1} dt' \frac{H_I(t')}{\hbar}\nonumber\\
&&+(-i)^2\int_{t_0}^{t_1} dt' \int_{t_0}^{t'} dt'' \frac{H_I(t')H_I(t'')}{\hbar^2}+\ldots \nonumber .
\end{eqnarray}
Often after time-ordering some of the terms appearing, depicted as Feynman diagrams, can be grouped into classes and summed. Here instead we use the Magnus expansion \cite{magnus54,blanes09} in which a series expansion is developed as the argument of an exponential,
\begin{eqnarray}\label{magnusdef}
\uu(t_1,t_0) =\exp\big(&&\Omega_1(t_1,t_0)\\
&&+\Omega_2(t_1,t_0)+\Omega_3(t_1,t_0)+\ldots\big), \nonumber
\end{eqnarray}
where unless otherwise indicated we take $t_0 \to -\infty$ and $t_1 \to \infty $.
In the following we calculate the terms in such a series for a rather broad class of Hamiltonians relevant to nonlinear quantum optics.  We write write the time dependence of the interaction Hamiltonian as
\begin{eqnarray}\label{hamiltonian}
H_I(t)=\hbar \int d \boldsymbol{\omega} \left( h(\boldsymbol{ \omega}) e^{i \Delta t}+ h^\dagger(\boldsymbol{ \omega}) e^{-i \Delta t} \right),
\end{eqnarray}
where we use the shorthand notation
\begin{eqnarray}
h(\boldsymbol{ \omega})&=&h(\omega_1,\omega_2,\ldots,\omega_n),\\
d\boldsymbol{ \omega}&=&d \omega_1 d \omega_2 \ldots d\omega_n,\nonumber\\
\Delta&=& \Delta(\omega_1,\omega_2,\ldots,\omega_n).\nonumber
\end{eqnarray}
For the moment we leave the integration limits of the frequencies $\boldsymbol{\omega}$ unspecified; in the next sections we will specify them for particular cases.
\indent For the first order Magnus term one simply finds
\begin{eqnarray}\label{Omega1}
\Omega_1&=&-\frac{i}{\hbar} \int dt H_I(t)\\
&=&-2 \pi i \int d \boldsymbol{\omega} \left(h(\boldsymbol{ \omega}) +\hc \right) \delta(\Delta).
\end{eqnarray}
In the last equation we have adopted the convention that whenever the lower (or upper) limit of an integral over time is not specified it is meant to be understood as $-\infty$ (or $+\infty$). The term \ref{Omega1}, exponentiated, corresponds to the sum of the Taylor-like series for $\uu$ that could be constructed by time-ordering the series (\ref{dysondef}) and ignoring the fact that the Hamiltonians $H_I(t)$ at different times $t$ need not commute. In fact, if $[H_I(t),H_I(t')]=0$ for all $t$ and $t'$, then $\exp \Omega_1$ constitutes the \emph{exact} solution of the problem, since all the higher order Magnus terms vanish. Hence we refer to terms in the Magnus expansion expansion beyond $\Omega_1$ as ``time ordering corrections''.

In general no simplifications can be made to the higher order Magnus terms if no further assumptions are made about $H_I(t)$. We make the following:
\begin{eqnarray}\label{comm}
[ h(\boldsymbol{ \omega}), h(\boldsymbol{ \omega}')]=[ h^\dagger(\boldsymbol{ \omega}), h^\dagger(\boldsymbol{ \omega}')]=0.
\end{eqnarray}
which is indeed valid for the Hamiltonians usually employed to describe both type I and II parametric down conversion, four wave mixing, and frequency conversion, either if the pump is treated classically or if it is treated quantum mechanically.
Using property (\ref{comm}) and the fact that
\begin{eqnarray}\label{timeorder2}
\int dt \int ^{t} dt' e^{i \Delta t-i \Delta'
   t'}&=&2 \pi^2 \delta(\Delta) \delta(\Delta')\\
&&+2 \pi i \delta(\Delta-\Delta') \frac{\mathcal{P}}{ \Delta} , \nonumber
\end{eqnarray}
with $\mathcal{P}$ indicating a principal value, the second order Magnus term, generally given by
\begin{eqnarray}
\Omega_2=\frac{(-i)^2}{2 \hbar^2}\int  dt \int ^{t} dt' \left[H_I(t),H_I(t') \right],
\end{eqnarray}
can be simplified to
\begin{eqnarray}\label{magnus2}
\Omega_2=-2 \pi i \fint d \boldsymbol{\omega} d \boldsymbol{\omega}' 
\left[ h(\boldsymbol{\omega}), h^\dagger(\boldsymbol{\omega}')  \right] \frac{\delta(\Delta-\Delta')}{\Delta}.
\end{eqnarray}
In the last equation we have used the principal value integral symbol $\fint$. This is necessary because the result (\ref{timeorder2}) is only valid under a principal value integration sign. Only the second term of (\ref{timeorder2}) contributes to the second order term. This is to be expected since the first term is just 
\begin{eqnarray}\label{to2}
2 \pi^2 \delta(\Delta)\delta(\Delta') = \frac{1}{2} \int dt \int dt' e^{i \Delta t-i \Delta'   t'},
\end{eqnarray}
 which is the expression that would be obtained from (\ref{timeorder2}) ignoring the time ordering; note the difference in limits between Eq. (\ref{to2}) and Eq. (\ref{timeorder2}).

Now let us look at the third order Magnus term, which in general
\begin{align}
&\Omega_3=\frac{(-i)^3}{6 \hbar^3}\int dt \int ^{t} dt' \int ^{t'} dt'' \times \\
&\left( \left[H_I(t),\left[H_I(t'),H_I(t'') \right] \right]+\left[\left[H_I(t),H_I(t') \right],H_I(t'') \right] \right). \nonumber
\end{align}
Here the relevant time integrals take the form
\begin{eqnarray}\label{triple}
\int &dt& \int ^{t} dt' \int ^{t'} dt'' e^{i \mu t+i \nu t'+i \xi t''}\\
&=&2 \pi ^3 \delta \left(\mu \right) \delta \left(\nu\right) \delta \left(\xi\right)\nonumber\\
&&-2  \pi ^2 i \delta \left(\nu+\xi\right) \delta \left(\mu+\nu+\xi\right)\frac{\mathcal{P}}{\xi}\nonumber\\
&&+2  \pi ^2 i \delta \left(\mu+\nu\right) \delta \left(\mu+\nu+\xi\right)\frac{\mathcal{P}}{\mu}\nonumber\\
&&-2 \pi  \delta \left(\mu+\nu+\xi\right)
\frac{\mathcal{P}}{\mu+\nu} \frac{\mathcal{P}}{\mu}.\nonumber
\end{eqnarray}
Because of the condition (\ref{comm}) the only term that contributes to $\Omega_3$ in (\ref{triple}) is the last one, and then upon relabeling one obtains:
\begin{eqnarray}\label{mag3}
\Omega_3=-2 \pi i \fint && d \boldsymbol{\omega} d \boldsymbol{\omega}' d \boldsymbol{\omega}'' \left([h(\boldsymbol{\omega}),[h(\boldsymbol{\omega}'), h^\dagger(\boldsymbol{\omega}'')]]+\hc \right)\nonumber\\
&& \times \frac{ \delta \left(\Delta+\Delta'-\Delta''\right)}{3} \left(\frac{1}{\Delta \Delta''}+\frac{1}{\Delta \Delta'} \right).
\end{eqnarray}
We emphasize that the simple form obtained for $\Omega_3$ critically depends on assumption (\ref{comm}). If this did not hold, then the second and third terms in (\ref{triple}) would also contribute, leading to a much more complicated expression. However, even if (\ref{comm}) were not to hold the first term of (\ref{triple}) would never contribute to $\Omega_3$, which is related to the fact that it is proportional to terms that would arise if time order were ignored, and their effects have all been included in $\Omega_1$.

Finally, let us note that the Magnus expansion is more robust in preserving qualitative features of the exact solution for $\uu$ than the Dyson expansion, particularly for problems in nonlinear quantum optics. Generally, of course, the Magnus expansion truncated at any level will always give rise to a unitary evolution operator if the Hamiltonian is Hermitian, as is assumed when writing (\ref{hamiltonian}), while a truncated Dyson expansion will not. More specifically, the Magnus expansion truncated at any level will always give rise, in the Heisenberg picture, to a linear Bogoliubov transformation of the bosonic operators if the Hamiltonian is quadratic in them. Since such quadratic bosonic Hamiltonians (QBH) often appear in nonlinear quantum optics, and since for such a Hamiltonian the exact solution for the Heisenberg picture evolution of the bosonic operators indeed takes the form of a linear Bogoliubov transformation \cite{Braunstein05}, this is a very desirable feature of the Magnus approximation. Here is the proof that any Magnus approximation displays this feature: From the general commutator identity
\begin{eqnarray}
[A B, C D]&=&[A,C]BD+A[B,C]D\\
&&+C[A,D]B+CA[B,D]. \nonumber
\end{eqnarray}
we see that if $A,B,C,D$ are all bosonic operators then the commutators on the right hand side are all $c-$numbers, and in fact if they do not vanish they are delta functions, either Kronecker or Dirac. Since all the terms $\Omega_1, \Omega_2$ etc., in the Magnus series involve commutators, it is clear that for a QBH all the terms $\Omega_i$ will be quadratic functions of the bosonic operators, and the $M^{\text{th}}$ order approximation to the Magnus expansion
\begin{eqnarray}\label{orderm}
\uu_M(t_f,t_0)&=&\exp(\Omega_1(t_f,t_0)+\Omega_2(t_f,t_0)+\nonumber\\ 
&&\ldots+\Omega_M(t_f,t_0))\nonumber\\
&=&\exp(\mathcal{O}_M(t_f,t_0)), 
\end{eqnarray}
where we reinstate general initial and final times, $t_0$ and $t_f$ respectively, is characterized by an operator $\mathcal{O}(t_f,t_0)$ that is anti-Hermitian and a quadratic function of the boson operators. Now in the Heisenberg picture and initial operator $A(t_0)$ evolves to a later operator $A(t)$ according to
\begin{eqnarray}
A(t_f)&=&\uu^\dagger(t_f,t_0) A(t_0) \uu(t_f,t_0).
\end{eqnarray}
Using the approximation given by (\ref{orderm}) and the Baker-Campbell-Hausdorff (BCH) formula 
\begin{eqnarray}\label{BCH}
e^{X}Y e^{-X} &=&Y+\left[X,Y\right]+\frac{1}{2!}[X,[X,Y]]+\cdots ,
\end{eqnarray}
 we find
\begin{eqnarray}\label{transinout}
A(t_f)&=&A(t_0)+[A(t_0),\mathcal{O}_M(t_f,t_0)]\\
&&+\frac{1}{2!}[[A(t_0),\mathcal{O}_M(t_f,t_0)],\mathcal{O}_M(t_f,t_0)]+\ldots \nonumber.
\end{eqnarray}
Using the general commutator identity
\begin{eqnarray}
[A,BC]=[A,B]C+B[A,C],
\end{eqnarray}
we see that if $A(t_0)$ is a bosonic operator the commutators on the right hand side of (\ref{transinout}) will all be bosonic operators, and thus the Heisenberg bosonic operators at $t_f$ are indeed  linear functions of the Heisenberg bosonic operators at $t_0$. This completes the proof. In the Schr\"{o}dinger picture, the result implies that any Magnus approximation for evolution under a QBH will share the property of the exact solution that the general Gaussian nature of an initial state will be preserved; that is, the exact evolution constitutes a unitary Gaussian channel, and any Magnus approximation for the evolution constitutes a unitary Gaussian channel as well\cite{eisert05}.

\section{Time ordering in Spontaneous Parametric Down Conversion and Four Wave Mixing}\label{sec:pdc}
\noindent In type II SPDC an incoming photon at frequency $\omega_p$ is transformed into two photons of different polarizations at frequencies $\omega_a$ and $\omega_b$ which are roughly half the frequency of the incoming photon. Within the usual rotating wave and undepleted pump approximations, the interaction Hamiltonian governing this process in an effectively one-dimensional structure with all relevant modes propagating in the same direction, such as a channel or ridge waveguide, is (see Appendix \ref{chi2app} for a derivation)
\begin{eqnarray}\label{pdc}
H_I(t)=-\hbar \varepsilon  \int && d\omega_a d\omega_b d\omega_p e^{i \Delta t} \Phi(\omega_a,\omega_b,\omega_p) \nonumber\\
&&\times \alpha(\omega_p) a^\dagger(\omega_a) b^\dagger (\omega_b )+\hc, 
\end{eqnarray}
\begin{eqnarray}
\Delta=\omega_a+\omega_b-\omega_p.
\end{eqnarray}
where $a(\omega_a)$ and $b(\omega_b)$ are photon destruction operators for modes associated with the wave vectors $k_a(\omega_a)$ and $ k_b(\omega_b)$, constructed to satisfy the commutation relations
\begin{eqnarray}\label{commspdc}
[a(\omega),a^\dagger(\omega')]&&=[b(\omega),b^\dagger(\omega')]=\delta(\omega-\omega'),
\end{eqnarray}
and all other commutators between destruction and creation operators
vanishing; $\alpha \left( \omega _{p}\right) $ is a classical function
proportional to the electric field amplitude of the undepleted pump mode
with wave vector $k(\omega _{p})$.  The assumption (\ref{commspdc}) is well justified
because for type II SPDC the down converted photons appear in orthogonal
polarizations, and their frequency support is far away from the frequency
support of the pump mode characterized by $\alpha \left( \omega _{p}\right) $. Because of this the limits of integration for the frequencies of the modes will range from 0 to $\infty$. Nevertheless, since the bandwidth of the photons is much smaller than their central frequencies these integrals can be safely extended in their lower limit to $-\infty$.

The phase matching function $\Phi$ incorporates the spatial dependence of the modes and the nonlinearity profile; It takes the form
\begin{eqnarray}\label{pm1}
\Phi(\omega_a,\omega_b,\omega_p) = \sinc \left(\frac{\Delta k L}{2} \right),\\
\Delta k =k_a(\omega_a)+k_b(\omega_b)-k_p(\omega_p),
\end{eqnarray}
when the nonlinearity is assumed to be uniform in a region of space from $-L/2$ to $L/2$, and in Eq. (\ref{pdc}) $\varepsilon$ is a dimensionless constant incorporating parameters such as the strength of the nonlinearity profile and the effective cross sectional area of the crystal. 
There will be corrections to the integrand of (\ref{pdc}) depending in a benign way on the frequencies \cite{grice97}, and for simplicity we neglect them here.\\

In SFWM a pair of photons from a pump mode at frequency $\omega _{p}$ are
converted into a pair of photons of different frequencies $\omega _{p}\pm
\Delta \omega _{p}$.  The description of this process is inherently more
complicated than that of SPDC, but we show in Appendix C that if the pump
field is treated classically and undepleted, then the Hamiltonian governing the process is formally identical to (\ref{pdc}) if the effects of group velocity dispersion on the propagation of the pump pulse can be neglected. \ In that limit we find 
\begin{eqnarray}
H_{I}(t)=-\hbar \varepsilon \int &&  d\omega _{a}d\omega _{b}d\omega _{+}e^{i\tilde{\Delta}t}\tilde{\Phi}(\omega _{a},\omega _{b},\omega _{+}) \nonumber\\
&&\times \tilde{\alpha}(\omega _{+})a^{\dagger }(\omega _{a})b^{\dag }(\omega _{b})+\hc,
\label{HISFWM}
\end{eqnarray}
where here $\tilde{\Delta}=\omega _{a}+\omega _{b}-\omega _{+}$, and 
\begin{eqnarray}
\tilde{\Phi}(\omega _{a},\omega _{b},\omega _{+}) =\text{sinc}\left( \frac{\Delta k L}{2}\right) , \\
\Delta k = k_{a}(\omega _{a})+k_{b}(\omega _{b})-2k_{p}\left( \frac{\omega_{+}}{2}\right) ,
\end{eqnarray}
and now $\tilde{\alpha}(\omega _{+})$ is proportional to the square of the
amplitude of the electric field of the pump.  Again we have assumed that
the generated photons at $\omega _{a}$ and $\omega _{b}$ are characterized
by distinct polarizations and/or frequency ranges -- the latter, for
example, might arise due to phase matching constraints \cite{moreno12} -- and thus the only nonvanishing commutation relations
are given by (\ref{commspdc}).

While the approach introduced in this paper can be applied to more
complicated scenarios for SFWM, we restrict ourselves here to situations
where (\ref{HISFWM}) is applicable; then the analysis of SPDC and SFWM is
essentially the same. \ The interaction Hamiltonian $H_{I}(t)$ takes the
form (\ref{hamiltonian}) with 
\begin{eqnarray}\label{Fdef}
h(\omega _{a},\omega _{b},\omega _{p}) &=&F(\omega _{a},\omega _{b},\omega_{p})a^{\dagger }(\omega _{a})b^{\dagger }(\omega _{b}), \\
F(\omega _{a},\omega _{b},\omega _{p}) &=&-\varepsilon \alpha (\omega_{p})\Phi (\omega _{a},\omega _{b},\omega _{p}),
\end{eqnarray}%
for SPDC; for SFWM we simply replace $\omega _{p}$ by $\omega _{+}$, $\alpha
(\omega _{p})$ by $\tilde{\alpha}(\omega _{+})$, and $\Phi (\omega
_{a},\omega _{b},\omega _{c})$ by $\tilde{\Phi}(\omega _{a},\omega
_{b},\omega _{+})$. \ So although in the calculations below we explicitly
use the notation for SPDC, the generalization to SFWM is trivial.

\begin{figure}
\centering
\includegraphics[width=0.40\textwidth]{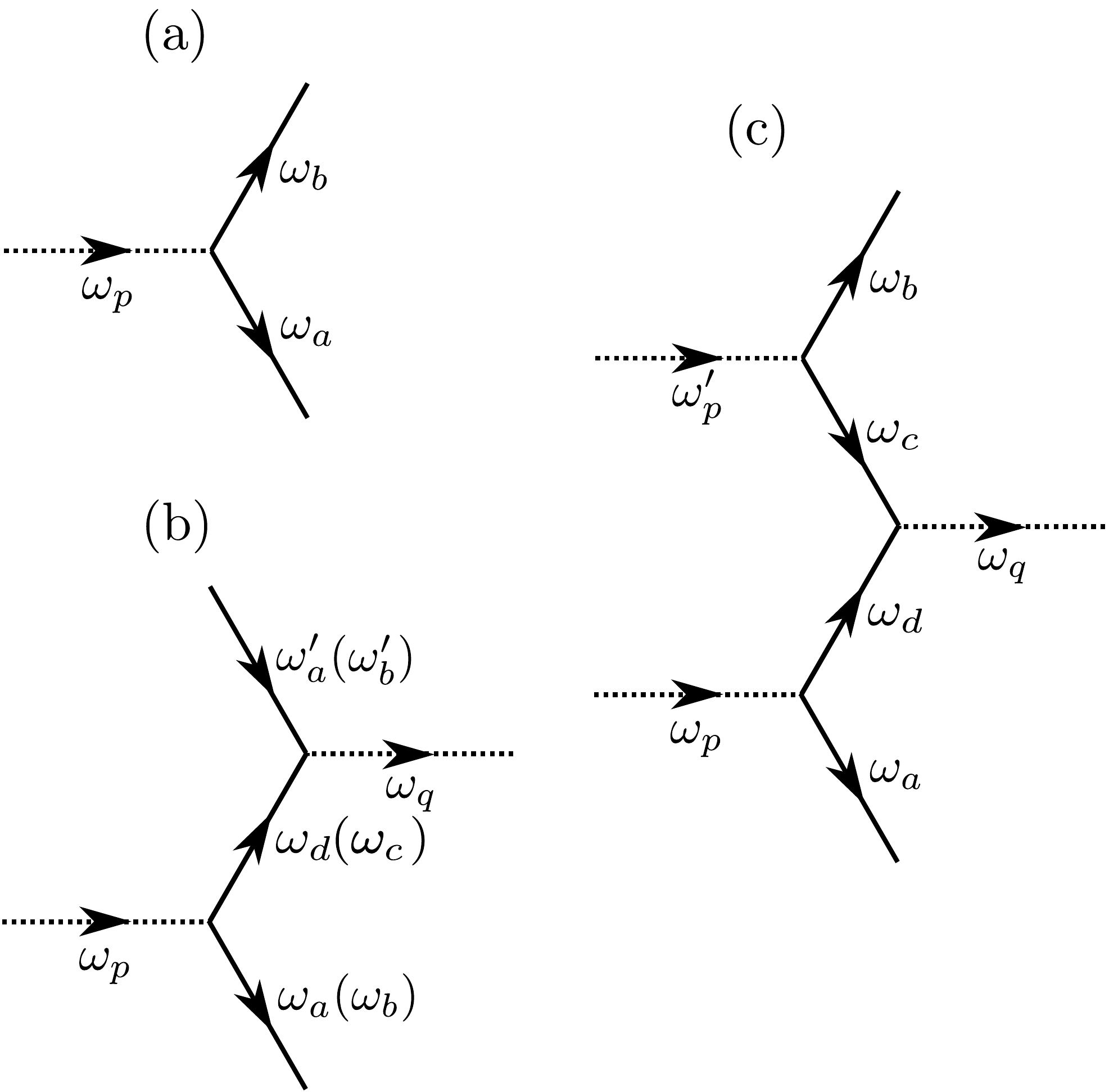}
\caption{\label{fdiag} Diagrams representing the first, second and third order Magnus terms for SPDC. Dashed lines are used to represent pump photons, full lines are used to represent lower energy down converted ones. (a) depicts a photon of frequency $\omega_p$ being converted to two photons of frequencies $\omega_a$ and $\omega_b$. (b) depicts the second order Magnus term in which one of the photons from a down-conversion event is, with the help of a low energy photon previously present, up converted to a pump photon. Finally, (c) depicts the third order Magnus term in which a pair of photons from two previous down-conversion events is converted to a pump photon.
}
\end{figure}

With the notation introduced in the paragraphs above, the first order Magnus term is
\begin{eqnarray}
\Omega_1=-2 \pi i \int d\omega_a d\omega_b  \big( && J_1(\omega_a,\omega_b)a^\dagger(\omega_a) b^\dagger (\omega_b)\\
&&+ \hc  \big), \nonumber
\end{eqnarray}
with
\begin{eqnarray}\label{J1pdc}
J_1(\omega_a,\omega_b)&=&F(\omega_a,\omega_b,\omega_a+\omega_b).
\end{eqnarray}
This process can be represented by the diagram (a) in Fig. \ref{fdiag} in which a photon of frequency of $\omega_p$ (or two photons of frequency $\omega_p$ for SFWM) is converted to two photons of frequency $\omega_a$ and $\omega_b$; we associate it with the term $  J_1\left(\omega _a,\omega _b\right) a^\dagger (\omega _a)b^\dagger (\omega _b)$. To have a Hermitian Hamiltonian the reverse process needs to be possible and this is associated with $  \bar{J}_1\left(\omega _a,\omega _b\right) a (\omega _a) b (\omega _b)$ where $\bar{x}$ is the complex conjugate of $x$.

We note that with the approximation of keeping only the first term in the Magnus expansion, as well as the assumption that its effect is small, we have
\begin{eqnarray}
\ket{\psi(\infty)}&=&\uu \ket{\psi(-\infty)}\approx  e^{\Omega_1}\ket{\psi(-\infty)}\nonumber \\
&=&(\mathbb{I}+\Omega_1) \ket{\psi(-\infty)} =\mathbb{I}+T_1 \ket{\psi(-\infty)},
\end{eqnarray}
where $T_1$ is the first term of the Dyson series (\ref{dysondef}), this is, as shown in the last equation, identical to the first order Magnus term $\Omega_1$.
For the second order Magnus term we need the following:
\begin{eqnarray}
\left[ h(\boldsymbol{\omega}), h^\dagger (\boldsymbol{\omega}') \right]=&-&\delta(\omega _b'-\omega _b)F\left(\omega _a,\omega _b,\omega _p\right) \times \\
&&\bar{F}\left(\omega _a',\omega _b',\omega _p'\right) a ^{\dagger } (\omega _a)a(\omega _a')\nonumber\\
&-&\delta(\omega _a'-\omega _a)F\left(\omega _a,\omega _b,\omega _p\right) \times \nonumber\\
&&\bar{F}\left(\omega _a',\omega _b',\omega _p'\right) b ^{\dagger }(\omega _b)b(\omega _b')\nonumber.
\end{eqnarray}
After some relabeling the second order Magnus term is
\begin{eqnarray}\label{fun2}
\Omega_2=-2 \pi i \Big( \int && d\omega_a d\omega_a'  G_2^a(\omega_a ,\omega_a') a ^{\dagger}\left(\omega_a\right)a\left(\omega_a'\right)  \\
&&+\int  d\omega_b d\omega_b'  G_2^b(\omega_b ,\omega_b') b^{\dagger}\left(\omega_b\right) b\left(\omega_b'\right)\Big),\nonumber
\end{eqnarray}
with
\begin{eqnarray}\label{g2}
G_2^a(\omega_a&,&\omega_a') =
  \int   d\omega_{d} \fint  \frac{d\omega_p}{\omega_p} \times \\
&& F\left(\omega_a,\omega _d,\omega _a+\omega _p+ \omega _d \right) \bar{F}\left(\omega _a',\omega _d,\omega _a'+\omega _p+\omega _d\right), \nonumber
\end{eqnarray}
\begin{eqnarray}\label{h2}
G_2^b(\omega_b&,&\omega_b')=  \int   d\omega_c \fint \frac{d\omega_p}{\omega_p} \times\\
&& F\left(\omega _c,\omega _b,\omega _b+\omega_c+\omega _p\right) \bar{F}\left(\omega _c,\omega _b',\omega  _b'+\omega _p+\omega_c\right). \nonumber
\end{eqnarray}
The last expression shows that the second order term corresponds to a frequency conversion process in which a photon of frequency $\omega_a'$ ($\omega_b'$) is converted to one of frequency $\omega_a$ ($\omega_b$) with probability amplitude $G_2^a (G_2^b)$. If one recognizes that $F$ is associated with down conversion and $\bar{F}$ with up conversion, then Eq. \ref{g2} (\ref{h2}) can be read as follows: first a photon at the pump energy (or two photons for the case of SFWM) is down converted to two photons of energies $\omega_a$ ($\omega_b$) and $\omega_d (\omega_c)$. Later the photon of energy $\omega_d (\omega_c)$ interacts with a  photon of energy $\omega_a'$ ($\omega_b'$) to be up converted to another photon of energy close to the pump energy (or a pair of photons for SFWM), which effectively gives a frequency conversion process from $\omega_a'$($\omega_b'$) to $\omega_a$ ($\omega_b$). This is sketched in diagram (b) in Fig. \ref{fdiag}.

For the third order Magnus term we need to calculate the following:
\begin{eqnarray}\label{mag3p}
[h(\boldsymbol{\omega})&,&[h(\boldsymbol{\omega}'), h^\dagger(\boldsymbol{\omega}'')]]=\\
&& F\left(\omega _a,\omega _b,\omega _p\right) F\left(\omega _a',\omega _b',\omega _p'\right) \bar{F}\left(\omega _a'',\omega _b'',\omega _p''\right)\times \nonumber\\
&&   \Big(\delta \left(\omega _a''-\omega _a'\right) \delta \left(\omega _b''-\omega _b\right)
   b^{\dagger }\left(\omega _b'\right)a^{\dagger }\left(\omega _a\right)\nonumber\\
&&
+\delta
   \left(\omega _a''-\omega _a\right) \delta \left(\omega _b''-\omega _b'\right) b^{\dagger }\left(\omega
   _b\right)a^{\dagger }\left(\omega _a'\right)\Big). \nonumber
\end{eqnarray}
Using (\ref{mag3}), after some rearranging of the dummy integration variables we obtain 
\begin{eqnarray}
\Omega_3&&=-2 \pi i \int d\omega_a d\omega_b a^\dagger(\omega_a) b^\dagger(\omega_b) \\
&&\fint d\omega_c d\omega_d d\omega_p d\omega_q \frac{   \bar{F}\left(\omega _d,\omega _c,-\omega _a-\omega _b+\omega _p+\omega _q\right)}{3}\nonumber\\
&&\times F\left(\omega _a,\omega _c,\omega _p\right) F\left(\omega _d,\omega _b,\omega _q\right) \Big\{\nonumber\\
&&\frac{2}{\left(\omega _a+\omega _c-\omega _p\right) \left(\omega _b+\omega _d-\omega _q\right)}\nonumber\\
&&+ \frac{1}{\left(\omega _a+\omega _c-\omega _p\right) \left(\omega _a+\omega _b+\omega _c+\omega_d-\omega _p-\omega _q\right)} \nonumber\\
&& +\frac{1}{\left(\omega _b+\omega _d-\omega _q\right) \left(\omega _a+\omega _b+\omega _c+\omega_d-\omega _p-\omega _q\right)} \nonumber
\Big\}\nonumber.
\end{eqnarray}
Switching $\omega_p \to \omega_p+\omega_a+\omega_c$ and $\omega_q \to \omega_q+\omega_b+\omega_d$ the innermost integral becomes
\begin{eqnarray}\label{bpv}
&&\int  d\omega_c d\omega_d \fint d\omega_p d\omega_q \frac{ \bar{F}\left(\omega _d,\omega _c,\omega _c+\omega _d+\omega_p+\omega _q\right)}{3} \nonumber\\
&&\times F\left(\omega _a,\omega _c,\omega _a+\omega _c+\omega _p\right) F\left(\omega _d,\omega _b,\omega _b+\omega _d+\omega _q\right)\nonumber\\
&&\times \Big\{\frac{2}{\omega _p \omega _q}+\frac{1}{\omega_p+\omega_q}\left(\frac{1}{\omega_p}+\frac{1}{\omega_q} \right) \Big\}.
\end{eqnarray}
At this point we use the identity \cite{barnett02}
\begin{eqnarray}
\frac{\mathcal{P}}{\omega_p+\omega_q}\left(\frac{\mathcal{P}}{\omega_p}+ \frac{\mathcal{P}}{\omega_q} \right)= \frac{\mathcal{P}}{\omega_p} \frac{\mathcal{P}}{\omega_q}+\pi^2 \delta(\omega_p) \delta(\omega_q),
\end{eqnarray}
which is valid under the assumption that the test function over which the principal values are being calculated is well behaved. We can then write
\begin{eqnarray}
\Omega_3=-2 \pi i \int d\omega_a d\omega_b \big(&&J_3(\omega_a,\omega_b) a^\dagger(\omega_a)b^\dagger(\omega_b)\\
&&+\hc\big), \nonumber
\end{eqnarray}
with
\begin{eqnarray}\label{J_3}
&& J_3(\omega_a,\omega_b)=\int  d\omega_c d\omega_d \Big\{ \frac{\pi^2}{3} 
\bar{F}\left(\omega _c,\omega _d,\omega _c+\omega _d\right)   \times \\
&& F\left(\omega _a,\omega _d,\omega _a+\omega _d\right) F\left(\omega _c,\omega _b,\omega _b+\omega_c\right) \nonumber\\
&&+\fint  \frac{d \omega_p}{\omega_p} \frac{d \omega_q}{\omega_q}  \bar{F}\left(\omega _c,\omega _d,\omega _c+\omega _d+\omega _p+\omega _q\right) \times \nonumber\\
&& F\left(\omega _a,\omega _d,\omega _a+\omega _d+\omega _p\right) F\left(\omega _c,\omega _b,\omega _b+\omega_c+\omega _q\right)  \Big\}. \nonumber
\end{eqnarray}
As before, a diagram representing the third order processes can be drawn. This corresponds to two down conversion processes in which one of the converted photons from each process is later used for up-conversion, as is seen in diagram (c) of Fig. \ref{fdiag}. 

Note that the treatment for type I SPDC follows easily from what was done above.  One needs only to replace $b$ by $a$ to characterize the fact that
both down converted photons are generated with the same polarization and in the same frequency range; the quantity $F$ in (\ref{Fdef}) will be symmetric in the first two arguments, $F(x,y,z)=F(y,x,z)$, and the expression for $\varepsilon $ will differ by a factor of two (see Appendix B).

Finally, let us mention that we have developed a numerical library \cite{github} that uses the \verb|cubature| software package \cite{cubature} to efficiently calculate the different Magnus terms for SPDC and SFWM within the approximations introduced in this paper.

\section{Time ordering in Frequency Conversion}\label{sec:fc}
\noindent Frequency conversion (FC) is another nonlinear process in which three fields interact. In this section we explicitly consider FC employing a $\chi_2$ nonlinearity, but the extension to treat FC by a $\chi_3$ nonlinearity follows in the same way the description of photon generation by SFWM followed from the description of photon generation by SPDC/ For a $\chi_2$ nonlinearity, FC occurs when a pump photon with frequency $\omega_p$ fuses with a photon with frequency $\omega_a$ to create a higher energy photon with frequency $\omega_b=\omega_a+\omega_p$. The Hamiltonian governing this process is derived in a completely analogous manner to the way (\ref{pdc}) is derived in Appendix \ref{chi2app}. The Hamiltonian governing the process is
\begin{eqnarray}\label{fc}
H_I(t)=-\hbar \varepsilon \int && d\omega_a d\omega_b d\omega_p  e^{i \Delta t} \Phi(\omega_a,\omega_b,\omega_p) \nonumber\\
&&\times \alpha(\omega_p) a(\omega_a) b^\dagger (\omega_b )+\hc,
\end{eqnarray}
where all the variables have the same meaning as in Eq. (\ref{pdc}) except that now the phase matching function for uniform non-linearity in the region $-L/2$ to $L/2$ is given by
\begin{eqnarray}\label{pm2}
\Phi(\omega_a,\omega_b,\omega_p) = \sinc\left(\frac{\Delta k L}{2} \right),\\
\Delta k =k_b(\omega_b)-k_a(\omega_a)-k_p(\omega_p),
\end{eqnarray}
and
\begin{eqnarray}
\Delta=\omega_b-\omega_a-\omega_p.
\end{eqnarray}
The interaction Hamiltonian (\ref{fc}) can be rewritten as
\begin{eqnarray}\label{notationfc}
H_I(t)&=& \int d\omega_a d\omega_b d\omega_p \left( e^{i \Delta  t} h(\omega_a,\omega_b,\omega_p)+ \hc  \right),
\end{eqnarray}
with
\begin{eqnarray}
h(\omega_a,\omega_b,\omega_p)=a (\omega _a)b^\dagger (\omega _b) F\left(\omega _a,\omega _b,\omega _p\right),
\end{eqnarray}
\begin{eqnarray}\label{f2}
F(\omega_a,\omega_b,\omega_p)=-\varepsilon \alpha(\omega_p)  \Phi(\omega_a,\omega_b,\omega_p).
\end{eqnarray}
In this case the first Magnus term is
\begin{eqnarray}\label{Omega1fc}
\Omega_1=- 2 \pi i\int d\omega_a d\omega_b \big(&& J_1(\omega_a,\omega_b)a(\omega_a) b^\dagger (\omega_b)\\
&&+ \hc  \big). \nonumber
\end{eqnarray}
with
\begin{eqnarray}\label{magnus1fc}
J_1(\omega_a,\omega_b)&=&F(\omega_a,\omega_b,\omega_b-\omega_a).
\end{eqnarray}
For the higher order terms we find:
\begin{eqnarray}\label{fun2fc}
\Omega_2=-2 \pi i \Big( \int && d\omega_a d\omega_a'  G_2^a(\omega_a ,\omega_a') a ^{\dagger}\left(\omega_a\right)a\left(\omega_a'\right)  \\
&&+\int  d\omega_b d\omega_b'  G_2^b(\omega_b ,\omega_b') b^{\dagger}\left(\omega_b\right) b\left(\omega_b'\right)\Big)\nonumber.
\end{eqnarray}
with
\begin{eqnarray}\label{g2fcn}
G_2^a(\omega_a&,&\omega_a') =
 -\int   d\omega_{q} \fint  \frac{d\omega_p}{\omega_p} \times \\
&& F\left(\omega_a',\omega _q,\omega _q-\omega _a'+\omega _p \right) \bar{F}\left(\omega _a,\omega _q,\omega _q-\omega _a+\omega _p\right), \nonumber\\
G_2^b(\omega_b&,&\omega_b')=  \int   d\omega_q \fint \frac{d\omega_p}{\omega_p} \times\\
&& F\left(\omega _q,\omega _b, \omega _b-\omega_q+\omega _p\right) \bar{F}\left(\omega _q,\omega _b',\omega  _b'+\omega _p-\omega_q\right), \nonumber
\end{eqnarray}
and
\begin{eqnarray}
\Omega_3=-2 \pi i \int d\omega_a d\omega_b \left(J_3(\omega_a,\omega_b) a (\omega_a)b^\dagger(\omega_b)+\hc\right),
\end{eqnarray}
with
\begin{eqnarray}
&& J_3(\omega_a,\omega_b)=\int  d\omega_c d\omega_d \Big\{ \frac{-\pi^2}{3} 
\bar{F}\left(\omega _c,\omega _d,\omega _d-\omega _c\right)   \times \\
&& F\left(\omega _a,\omega _d,\omega _d-\omega _a\right) F\left(\omega _c,\omega _b,\omega _b-\omega_c\right) \nonumber\\
&&+\fint  \frac{d \omega_p}{\omega_p} \frac{d \omega_q}{\omega_q}  \bar{F}\left(\omega _c,\omega _d,\omega _d-\omega _c+\omega _p-\omega _q\right) \times \nonumber\\
&& F\left(\omega _a,\omega _d,\omega _d-\omega _a-\omega _q\right) F\left(\omega _c,\omega _b,\omega _b-\omega_c+\omega _p\right)  \Big\}. \nonumber
\end{eqnarray}
In the last equation it has been assumed again that the function $F$ is well behaved.

Note the very elegant connection the Magnus series provides between FC and SPDC. To obtain the hamiltonian of FC from that of SPDC it suffices to make the substitutions $a(\omega_a) \rightarrow a^\dagger(\omega_a), a^\dagger(\omega_a) \rightarrow a(\omega_a)$ and $\omega_a \rightarrow -\omega_a$. The corresponding Magnus terms of FC are obtained by performing exactly the same substitutions in the SPDC terms, and rearranging the sign of the variables in the principal value integrals over $\omega_q$ and $\omega_p$.

\section{Disentangling the Magnus expansion}\label{sec:dis}
\noindent The calculations for photon generation used the fact that the commutator of two two-mode squeezing Hamiltonians gives two single-mode frequency conversion terms,
\begin{eqnarray}
&&\left[ a^{\dagger }\left(\omega_p\right) b^{\dagger }\left(\omega _q\right)+\hc \ ,  a^{\dagger }\left(\omega _r\right) b^{\dagger }\left(\omega _s\right)+\hc \right]=\\
&& \delta(\omega _q-\omega _s) a^{\dagger }\left(\omega _r\right)a\left(\omega _p\right)+\delta(\omega _p-\omega _r)  b^{\dagger }\left(\omega_s\right)b\left(\omega _q\right)-\hc  \nonumber,
\end{eqnarray}
and the fact that the commutator of a single-mode frequency conversion term with a two-mode squeezing term is again a two-mode squeezing term,
\begin{eqnarray}
&&\left[ a^{\dagger }\left(\omega _s\right) a\left(\omega_r\right)+\hc \ ,  a^{\dagger } \left(\omega _p\right) b^{\dagger }\left(\omega_q\right)+\hc  \right]=\\
&&\delta(\omega _s-\omega _p) b^{\dagger} \left(\omega _q\right) a^{\dagger }\left(\omega _r\right)+ \delta(\omega _r-\omega_p) b^{\dagger }\left(\omega _q\right)a^{\dagger } \left(\omega _s\right)-\hc\nonumber
\end{eqnarray}
Likewise for FC they used the fact that the commutator of two two-mode frequency conversion terms gives single-mode frequency conversion terms,
\begin{eqnarray}
&&\left[ a\left(\omega_p\right) b^{\dagger }\left(\omega _q\right)+\hc \ ,  a\left(\omega _r\right) b^{\dagger }\left(\omega _s\right)+\hc \right]=\\
&& \delta(\omega _q-\omega _s) a^{\dagger }\left(\omega _p\right)a\left(\omega _r\right)+\delta(\omega _p-\omega _r)  b^{\dagger }\left(\omega_q\right)b\left(\omega _s\right)-\hc,  \nonumber
\end{eqnarray}
and that the commutator of a two-mode frequency conversion term and a single-mode frequency conversion term gives again a two-mode frequency conversion term,
\begin{eqnarray}
&&\left[ a^{\dagger }\left(\omega _s\right) a\left(\omega_r\right)+\hc ,  a^{\dagger } \left(\omega _p\right) b\left(\omega_q\right)+\hc  \right]=\\
&&\delta(\omega _s-\omega _p) a^{\dagger}\left(\omega _r\right) b \left(\omega _q\right) + \delta(\omega _r-\omega_p) a^{\dagger } \left(\omega _s\right) b\left(\omega _q\right)-\hc\nonumber
\end{eqnarray}

These results can be used to inductively show that the unitary that connects states at $t_0 \to  -\infty $ and $t \to \infty$ has to be of the form:
\begin{eqnarray}\label{magnusgen}
\mathcal{U}(t&,&t_0)=\exp\Big(-2 \pi i \int d \omega_1 d\omega_2 \\
&& \Big\{\tilde G^a(\omega_1,\omega_2)a^\dagger(\omega_1) a(\omega_2) +\tilde G^b(\omega_1,\omega_2)b^\dagger(\omega_1) b(\omega_2) \nonumber\\
&&+ (\tilde J(\omega_1,\omega_2)a^\dagger(\omega_1) b^\dagger(\omega_2) +\hc) \Big\} \Big) \nonumber,
\end{eqnarray}
with
\begin{eqnarray}
\tilde J=\sum_i J_{2i+1}; \quad \tilde G^a=\sum_i G^a_{2i}; \quad \tilde G^b=\sum_i G^b_{2i},
\end{eqnarray}
for SPDC and SFWM. For FC one only need to change 
\begin{eqnarray}\label{change}
a^\dagger(\omega_1) b^\dagger(\omega_2) \Rightarrow a^\dagger(\omega_1) b(\omega_2),
\end{eqnarray} 
in Eq. (\ref{magnusgen}).
The result (\ref{magnusgen}) confirms in detail the proof given at the end of section II. Eq. (\ref{magnusgen}) shows the great utility of the Magnus expansion. It provides us with the general form of the solution. One only needs to calculate the higher corrections to the coefficients that appear in front of the squeezing and frequency conversion terms to get an increasingly better approximation to the solution.

Diagrams similar to those presented in Fig. \ref{fdiag} can be drawn for the higher order terms. We emphasize that they are used here just as a convenient way of representing the terms in the Magnus expansion; they are \emph{not} Feynman diagrams, of course, and we have not provided any rule to actually do calculations using them. Yet they can help in understanding the physics. In particular, given that the amplitudes they represent will always be multiplied by only pairs of bosonic operators, no matter what the order of the diagram there can be only two free legs at the low down-converted frequencies, which correspond precisely to the frequencies of the bosonic operators that they multiply. The diagrams in Fig. \ref{fdiag} are precisely the three simplest diagrams satisfying this selection rule.

While the form (\ref{magnusdef}) of the unitary operator given by the Magnus expansion identifies the physics of the evolution, particularly as illustrated in the example of (\ref{magnusgen}), it cannot be directly used to calculate the evolution of an initial state, since the terms associated with the functions $J_{2i+1}$ do not commute with the terms associated with the functions $G_{2i}^a$ and $G_{2i}^b$. In this section we will provide a simple strategy to disentangle the Magnus expansion.We will be able to refactorize the unitary (\ref{magnusgen}) as a product of two unitaries. For the SPDC and SFWM  Hamiltonians (\ref{pdc},\ref{HISFWM}) these unitaries will correspond to a pure squeezing Hamiltonian and pure single mode frequency conversion
\begin{eqnarray}\label{fact}
\uu&=&\uu_{\text{sq}} \uu_{\text{fc}},\\
\uu_{\text{sq}}&=&e^{-2\pi i \int d \omega_a d \omega_b J(\omega_a,\omega_b) a^\dagger(\omega_a) b^\dagger(\omega_b)+\hc},\\
\uu_{\text{fc}}&=&e^{-2\pi i \int d \omega_a d \omega_a' G(\omega_a,\omega_a') a^\dagger(\omega_a) a(\omega_a')} \times\\
&&e^{-2\pi i \int d \omega_b d \omega_b' G(\omega_b,\omega_b') b^\dagger(\omega_b) b(\omega_b)}\nonumber.
\end{eqnarray}
The case of FC is completely analogous and one only need to follow prescription (\ref{change}).
The factorization (\ref{fact}) has two important features. The first one is that if time ordering is irrelevant then $J=J_1$ as given in Eq. (\ref{J1pdc}) and $G^a=G^b=0$. The second is that when acted on the vacuum the SPDC or SFWM state is simply
\begin{eqnarray}
\ket{\psi}&=&\uu \ket{\vac}=\uu_{\text{sq}}\uu_{\text{fc}}\ket{\vac}=\uu_{\text{sq}}\ket{\vac}\\
&=&e^{-2\pi i \int d \omega_a d \omega_b J(\omega_a,\omega_b) a^\dagger(\omega_a) b^\dagger(\omega_b)+\hc} \ket{\vac} \nonumber
\end{eqnarray}
that is $J$ is proportional to the joint spectral amplitude (JSA) of the down converted photons.

To obtain the factorization (\ref{fact}) we will use the BCH formula (\ref{BCH}) and the Zassenhaus formula \cite{casas12},
\begin{eqnarray}
e^{X+Y}&=& e^{X} e^{Y} e^{-\frac{1}{2} [X,Y]} e^{\frac{1}{6}(2[Y,[X,Y]]+ [X,[X,Y]] ) \cdots} .
\end{eqnarray}
The lowest order factorization that can be obtained by just going to second order in the Magnus expansion is to write
\begin{eqnarray}
\uu=e^{\Omega_1+\Omega_2}  \approx  e^{\Omega_1}e^{\Omega_2}=\uu_{\text{sq}} \uu_{\text{fc}}.
\end{eqnarray}
When applied to vacuum one simply obtains
\begin{eqnarray}
\ket{\psi_2}=\uu \ket{\vac}=e^{\Omega_1}\ket{\vac},
\end{eqnarray}
that is, up to second order the state is not modified by time ordering (since $\Omega_2\ket{\vac}=0$), reminiscent of what happens when the Dyson series is used for this same problem \cite{aggie11}. Nevertheless, in the Magnus expansion the meaning of the null contribution of the second order in SPDC and SFWM is transparent: this second order term requires at least one down-converted photon to be present and hence it will vanish when applied to the vacuum.\\
\noindent Thus it is clear that to see time-ordering corrections to the expression for the state of the generated photons it is necessary to go to at least to third order. In this case one finds
\begin{eqnarray}\label{spdc3}
e^{\Omega_1+\Omega_2+\Omega_3}&\approx& e^{\Omega_1+\Omega_3} e^{\Omega_2} e^{-\frac{1}{2}[\Omega_1+\Omega_3,\Omega_2]}\\
&\approx& e^{\Omega_1+\Omega_3} e^{\Omega_2} e^{-\frac{1}{2}[\Omega_1,\Omega_2]}\nonumber \\
&=& e^{\Omega_1+\Omega_3} e^{\Omega_2} e^{-\frac{1}{2}[\Omega_1,\Omega_2]}e^{-\Omega_2}e^{\Omega_2}\nonumber \\
&\approx&e^{\Omega_1+\Omega_3}  e^{-\frac{1}{2}e^{\Omega_2}[\Omega_1,\Omega_2]e^{-\Omega_2}}e^{\Omega_2}\nonumber \\
&\approx&e^{\Omega_1+\Omega_3}  e^{-\frac{1}{2}([\Omega_1,\Omega_2]+[\Omega_2,[\Omega_1,\Omega_2]])}e^{\Omega_2}\nonumber \\
&\approx&e^{\Omega_1+\Omega_3}  e^{-\frac{1}{2}[\Omega_1,\Omega_2]}e^{\Omega_2}\nonumber \\
&\approx&e^{\Omega_1+\Omega_3-\frac{1}{2}[\Omega_1,\Omega_2]}e^{\Omega_2}. \nonumber
\end{eqnarray}
Here we ignore any fourth order term, resolved the identity as $\mathbb{I}=e^{-\Omega_2}e^{\Omega_2}$, used the BCH and Zassenhaus identities and the following relation $e^W e^V e^{-W}\equiv e^{e^W V e^{-W}} $.
From the last derivation is seen that to calculate the first non-vanishing correction to the JSA we not only need $\Omega_3$ but also $[\Omega_1,\Omega_2]$. This extra third order correction is easy to understand: The diagram corresponding to $\Omega_2$ needs an incoming photon, which can only be provided by a previous down conversion process event included in $\Omega_1$. This process will be of course of third order since it includes two down conversion events and one up conversion event. With $\Omega_1$ a two mode squeezing operator and $\Omega_2$ containing two single mode frequency converters, their commutator is another two mode squeezing operator:
\begin{eqnarray}
\frac{[\Omega_1, \Omega_2 ]}{2}=-2 \pi && i^2 \int d \omega_a d \omega_b \times \\
&&(K_3(\omega_a,\omega_b) a^\dagger(\omega_a) b^\dagger(\omega_b) +\hc) \nonumber,
\end{eqnarray}
with
\begin{eqnarray}
K_3(\omega_a,\omega_b)&=& \pi  \int d \omega \Big(G_2^a\left(\omega _a,\omega \right) J_1\left(\omega ,\omega _b\right)\\
&&+J_1\left(\omega _a,\omega \right) G_2^b\left(\omega _b,\omega \right)  \Big)\nonumber.
\end{eqnarray}
Eq. (\ref{spdc3}) provides a compact formula accounting for the effects of time ordering, and leads to an identification of the form of the joint spectral amplitude, which to this level of approximation is proportional to
\begin{eqnarray}
J(\omega_a,\omega_b)&=&J_1(\omega_a,\omega_b)\\
&&+\left(J_3(\omega_a,\omega_b)-i K_3(\omega_a,\omega_b)\right) \nonumber,
\end{eqnarray}
where the first contribution $J_1(\omega_a,\omega_b)$ neglects effects of time-ordering, while the additional terms take corrections into account to all orders lower than the fifth (this last fact we shall prove shortly). This illustrates that as the intensity of the pump field increases and the corrections become larger not only are more squeezed photons generated in the same modes, but their spectral properties change. In particular, if the functions characterizing the pump pulse and the phase-matching functions are real, then as the pump intensity is increased the phase structure of the joint spectral amplitude evolves from trivial to nontrivial.

From the general properties of the Magnus expansion, \emph{i.e.} the fact that the $n^{\text{th}}$ order Magnus terms is made of $n-1$ commutators, we know that all even Magnus terms will be pairs of single mode frequency conversion generators, whereas all odd terms will be two mode squeezing operators. This fact can be used to show that the commutator of two Magnus terms $[\Omega_i,\Omega_j]$ (which is an $i+j$ order term) will be a frequency conversion term if $i+j$ is even and a squeezing term if $i+j$ is odd. With these two facts and after some algebra it can be shown, that the extension of (\ref{spdc3}) to fifth order is
\begin{eqnarray}
\uu= \exp(\Omega_1+\Omega_2+\Omega_3+\Omega_4+\Omega_5)= \uu_{\text{sq}} \uu_{\text{fc}},
\end{eqnarray}
where the squeezing and frequency conversion unitaries $\uu_{\text{sq}}$, $\uu_{\text{fc}}$ are
\begin{eqnarray}
\uu_{\text{sq}}&=&\exp(X),\\
X&=&\Omega_1+\Big(\Omega_3-\frac{1}{2}[\Omega_1,\Omega_2]\Big)+\Big(\Omega_5-\frac{1}{2}[\Omega_1,\Omega_4] \\
&&-\frac{1}{2}[\Omega_3,\Omega_2]+\frac{5}{6}[\Omega_2,[\Omega_2,\Omega_1]]+\frac{1}{2}[\Omega_1,[\Omega_1,\Omega_3]]\Big) \nonumber,\\
\uu_{\text{fc}}&=&\exp(Y),\\
Y&=&\Omega_2+\Big(\Omega_4+\frac{1}{2}[\Omega_1,[\Omega_1,\Omega_2]]\Big).
\end{eqnarray}

\indent We now consider difference frequency generation, where besides an injected pump pulse that is treated classically we also seed the modes at $\omega_a$ and/or $\omega_b$, with injected pulses described by coherent states. Here the output state can be written as
\begin{eqnarray}
\ket{\psi_3}&=&e^{\Omega_1+\Omega_3-\frac{1}{2}[\Omega_1,\Omega_2]}e^{\Omega_2} \times\\
&&e^{i \sum_{c={a,b}}\left(\int d \omega_c f_c(\omega_c) c^{\dagger}(\omega_c)+\hc\right)}\ket{\vac}\nonumber\\
&=& \left(e^{\Omega_1+\Omega_3-\frac{1}{2}[\Omega_1,\Omega_2]} \right)\times\nonumber\\
&&\left(e^{\Omega_2}e^{i \sum_{c={a,b}}\left(\int d \omega_c f_c(\omega_c) c^{\dagger}(\omega_c)+\hc\right)}e^{-\Omega_2}\right)\ket{\vac}.\nonumber
\end{eqnarray}
Keeping only terms up to third order, we find
\begin{eqnarray}\label{manexp}
&&e^{\Omega_2}e^{i \sum_{c={a,b}}\left(\int d \omega_c f_c(\omega_c) c^{\dagger}(\omega_c)+\hc\right)}e^{-\Omega_2}=\\
&&e^{i \sum_{c={a,b}}\left(\int d \omega_c f_c(\omega_c) e^{\Omega_2}c^{\dagger}(\omega_c)e^{-\Omega_2}+\hc\right)}=\nonumber\\
&&e^{i \sum_{c={a,b}}\left(\int d \omega_c f_c(\omega_c) \left(c^{\dagger}(\omega_c)+[\Omega_2,c^{\dagger}(\omega_c)] \right)+\hc\right)}\nonumber.
\end{eqnarray}
The commutator inside the exponential can be easily calculated to be
\begin{eqnarray}
[\Omega_2,a^\dagger(\omega_a')]&=&-2 \pi i \int d \omega_a a^{\dagger}(\omega_a) G_2^a(\omega_a,\omega_a'),
\end{eqnarray}
\begin{eqnarray}
[\Omega_2,b^\dagger(\omega_b')]&=&-2 \pi i \int d \omega_b b^{\dagger}(\omega_b) G_2^b(\omega_b,\omega_b').
\end{eqnarray}
This shows that the first effect due to time ordering corrections in stimulated processes is to modify the shape of the injected seed pulses. 

Note that although the discussion in this section was phrased in terms of the SPDC and SFWM hamiltonians all its conclusions are directly applicable to the case of FC.

\section{Time ordering corrections in broadly phase-matched Processes}\label{sec:broad}
\noindent Here we consider time ordering corrections in the special limit that the phase matching function is independent of is variables,
\begin{eqnarray}\label{broad}
\Phi(\omega_a,\omega_b,\omega_p) \propto 1,
\end{eqnarray}
in Eqs. (\ref{pm1}) and (\ref{pm2}). This implies that the function $F$ in (\ref{Fdef}) and (\ref{f2}) is a function only of its last argument, $F(x,y,z)=F(z)$. 
We show below that if this holds the second and third order Magnus terms vanish. We conjecture that in this limit \emph{all} Magnus terms $\Omega_i$  beyond $\Omega_1$ vanish for PDC and FC; in the special case of FC in which the initial state is a single photon this has been shown to hold\cite{donohue14}. \\
\indent To proceed, we look at $G_2^a(\omega_a, \omega_a')$ as a typical term. For PDC, in the limit (\ref{broad}) it becomes
\begin{eqnarray}
G_2^a(\omega_a&,&\omega_a') =
  \int   d\omega_{q} \fint  \frac{d\omega_p}{\omega_p} \times \\
&& F\left(\omega_a,\omega _q,\omega _a+\omega _p+ \omega _q \right) \bar{F}\left(\omega _a',\omega _q,\omega _a'+\omega _p+\omega _q\right) \nonumber\\
&&= \int   d\omega_{q} \fint  \frac{d\omega_p}{\omega_p} F\left(\omega _a+\omega _p+ \omega _q \right) \bar{F}\left(\omega _a'+\omega _p+\omega _q\right). \nonumber
\end{eqnarray}
In the last equation we are free to move the origin of the integral in $\omega_q$ by $\omega_p$ to get:
\begin{eqnarray}
G_2^a = \int   d\omega_{q}  F\left(\omega _a+ \omega _q \right) \bar{F}\left(\omega _a'+\omega _q\right) \fint  \frac{d\omega_p}{\omega_p},
\end{eqnarray}
using the identity
\begin{eqnarray}\label{id1}
 \fint \frac{dx}{ x+y}= 0,
\end{eqnarray}
the innermost integral is identically zero.
The same kind of argument shows that $G_2^b$ vanishes. Finally, to show that the third order term is also zero we proceed along similar lines. Using expression (\ref{bpv}), with  $F(x,y,z)=F(z)$ we obtain:
\begin{eqnarray}
&&\int  d\omega_c d\omega_d \fint d\omega_p d\omega_q \frac{
\bar{F}\left(\omega _c+\omega _d+\omega_p+\omega _q\right)
}{3} \nonumber\\
&&\times F\left(\omega _a+\omega _c+\omega _p\right) F\left(\omega _b+\omega _d+\omega _q\right)\nonumber\\
&&\times \Big\{\frac{2}{\omega _p \omega _q}+\frac{1}{\omega_p+\omega_q}\left(\frac{1}{\omega_p}+\frac{1}{\omega_q} \right) \Big\}.
\end{eqnarray}
and as before we shift the integration axes, in this case, $\omega_c \rightarrow \omega_c-\omega_q$, $\omega_d \rightarrow \omega_d-\omega_p$ to get:
\begin{eqnarray}
\int  d\omega_c d\omega_d &&  \bar{F}\left(\omega _c+\omega _d\right) F\left(\omega _a+\omega _d\right) F\left(\omega _b+\omega_c\right) \times  \\
&& \fint d \omega_p d \omega_q \left\{ \frac{2}{\omega_p \omega_q}+\frac{1}{\omega_p+\omega_q}\left(\frac{1}{\omega_p}+\frac{1}{\omega_q} \right) \right\}= 0. \nonumber
\end{eqnarray}

\section{Conclusions}\label{sec:end}
\noindent In this paper we have used the Magnus expansion to construct solutions to the time dependent Schr\"{o}dinger equation describing three nonlinear processes governed by Hamiltonians that do not commute at different times: spontaneous parametric down-conversion, spontaneous four wave mixing, and frequency conversion. In the derivation of the Magnus terms for these three processes we found a rather simple form for the second and third order Magnus terms for a broad class of Hamiltonians that satisfy Eq. (\ref{comm}). The Magnus expansion solution that we found for the time evolution operator was shown to be Gaussian preserving; equivalently, when the unitary operator is used to transform input and output bosonic operators the transformation is Bogoliubov linear. With the expansion we developed, it is easy to see why the second order effect due to the time ordering in SPDC vanishes: Essentially the second order process needs to be stimulated by a previous photon. We also showed how to calculate corrections to the joint spectral amplitude of the photons in SPDC and SFWM, and how even when the phase matching function and pump functions are real the JSA acquires a non-trivial phase structure due to time ordering; this structure can be investigated via tomographic methods\cite{sipe14}. We also explored how the effect of time ordering acts in stimulated PDC; in particular, we looked at the case when the lower energy photons are prepared in coherent states and showed that the first correction due to time ordering is to modify their spectral structure. Finally, the Magnus expansion also allowed us to argue that time ordering is irrelevant if the phase matching function is infinitely broad.

\begin{acknowledgments}
N.Q. gratefully acknowledges insightful conversations with J.M. Donohue and A.M. Bra{\'n}czyk and funding from NSERC Canada.
\end{acknowledgments}

\appendix

\section*{Appendix}
\noindent In these appendices we construct estimates for the form of the terms appearing in the interaction Hamiltonians we use in the text. \ If waveguides are employed in any actual experiment, a full description would need to take into account the form of the mode fields; if bulk crystals are employed instead, a full description would require the details of the focusing of the pump and radiation pattern of the generated photons. \ Since our purpose here is not to give a precise calculation for any particular system, but rather to identify the characteristic size and form of the terms needed in our calculations, we simply approximate the relevant field at each frequency in the linear regime as if it consisted of an area $A$ ``cut out'' of a linearly polarized plane wave propagating in a direction perpendicular to the plane of the area; this can be taken as a ``toy model'' of a waveguide mode.  In Appendix A we relate the energy in a pulse propagating in such a toy model of a waveguide to the field amplitude. With further approximations relating to the relevant tensor components of the nonlinear susceptibilites, in Appendix B we then construct the interaction Hamiltonian for a $\chi _{2}$ process, and in Appendix C we give the corresponding calculation for a $\chi_{3}$ process. 

\section{Energy in an Electromagnetic pulse}
\noindent A solution to Maxwell's equations in a medium with index of
refraction $n(\omega )$ is given by: 
\begin{eqnarray}
\mathbf{E}(z,t)&=&\left( \mathbf{\hat{x}}E_{0}e^{i(kz-\omega t)}\right) +\text{c.c.},\\
\mathbf{B}(z,t)&=&\left( \mathbf{\hat{y}}B_{0}e^{i(kz-\omega t)}\right) +\text{c.c.},
\end{eqnarray}%
with $E_{0}/B_{0}=c/n(\omega )$ and $k=k(\omega )$. To construct pulses we
take superpositions, 
\begin{eqnarray}\label{sup}
\mathbf{E}(z,t) &=&\mathbf{\hat{x}}\int_{0}^{\infty }d\omega E_{0}(\omega)e^{i(kz-\omega t)}+\text{c.c.}, \\
\mathbf{B}(z,t) &=&\mathbf{\hat{y}}\int_{0}^{\infty }d\omega B_{0}(\omega)e^{i(kz-\omega t)}+\text{c.c.} \nonumber
\end{eqnarray}%
In what follows we will assume that the amplitude functions $E(\omega )$ and 
$B(\omega )$ are very narrowly peaked around some central frequency $\nu $
and thus over the relevant integration range to good approximation we can
write $E(\omega )=E_{0}f(\omega )$ and $B(\omega )=B_{0}f(\omega ),$ where $%
E_{0}/B_{0}=c/n(\nu )$, taking $f(\omega )$ to vanish for $\omega <0$. With
this in mind, in a non-magnetic medium the Poynting vector is 
\begin{eqnarray}\label{poynt}
\mathbf{S}\left( z,t)\right) &=&\frac{\mathbf{E}(z,t)\;\mathbf{\times \;B(}z,t)%
}{\mu _{0}} \\
&=&\mathbf{\hat{z}}\frac{E_{0}B_{0}}{\mu _{0}}\int_{0}^{\infty }d\omega
d\omega ^{\prime } \times \\
&& \Big(f(\omega )f(\omega ^{\prime })e^{i(k(\omega
)+k(\omega ^{\prime }))z-(\omega +\omega ^{\prime })t}\nonumber \\
&&+f(\omega )\bar{f}(\omega ^{\prime })e^{i(k(\omega )-k(\omega ^{\prime }))z-(\omega -\omega^{\prime })t}\Big)  +\cc \nonumber
\end{eqnarray}%
where we use an overbar to indicate complex conjugation. \ Writing a
directed element of area $d\mathbf{A}$ as $\mathbf{\hat{z}}dA$, from
Poynting's theorem the energy carried in the pulse over an area $A$ is then
given by 
\begin{eqnarray}
U_{em} &=&\int_{-\infty }^{\infty }dt\int_{A}d\mathbf{A}\cdot \mathbf{S}(z,t)
\label{Uem} \\
&=&\frac{2\pi AE_{0}B_{0}}{\mu _{0}}\int_{0}^{\infty }d\omega \Big(f(\omega
)f(-\omega )e^{i(k(\omega )+k(-\omega ))z}  \notag \\
&&\quad \quad \quad \quad \quad \quad +|f(\omega )|^{2}+\text{c.c.}\Big).
\end{eqnarray}%
The first term is identically zero since $f(\omega )=0$ if $\omega <0$; if
we then, for example, approximate $f(\omega )$ for $\omega >0$ by 
\begin{eqnarray}
f(\omega ) &=&\frac{1}{\sqrt{2\pi }\sigma }\zeta (\omega -\nu ), \label{disp} \\
\zeta (\omega ) &=&\exp (-\omega ^{2}/(2\sigma ^{2})),  \notag
\end{eqnarray}%
and take 
\begin{equation}
k(\omega )=k(\nu )+(\omega -\nu )/v_{g},
\end{equation}%
with $v_{g}$ the group velocity at frequency $\nu $, we have 
\begin{eqnarray}
\mathbf{E}(z,t)&=&2\mathbf{\hat{x}}E_{0}e^{-\frac{1}{2}\sigma^{2}(t-z/v_{0})^{2}}\cos (k(\nu )z-\nu t), \\
\mathbf{B}(z,t)&=&2\mathbf{\hat{y}}B_{0}e^{-\frac{1}{2}\sigma^{2}(t-z/v_{0})^{2}}\cos (k(\nu )z-\nu t),
\end{eqnarray}%
where we have assumed that $\nu \gg \sigma $ and thus one can safely extend the lower integration limit in (\ref{sup}) from 0 to $-\infty $. \ Averaged over the period $2\pi /\nu $, the Poynting vector following from (\ref{poynt}) has a standard deviation $\sqrt{2}\sigma $, leading to a full width at half maximum (FWHM) of the intensity of the pulse given by $\text{FWHM}=2\sqrt{2\ln 2} (\sqrt{2}\sigma )$, or $\sigma =\text{FWHM}/4\sqrt{\ln 2}\approx 0.3\;\text{FWHM}$.  From (A6,A7), under the same approximations leading to (A10,A11) we have 
\begin{equation}
U_{em}=2A\epsilon _{0}n(\nu )cE_{0}^{2}\frac{\sqrt{\pi }}{\sigma }.
\end{equation}
This is a useful result that allows to relate the peak amplitude of the
electric field $E_{0}$ to the net energy in the pulse $U_{em}$: 
\begin{equation}\label{pulseE}
E_{0}=\sqrt{\frac{(U_{em}/\sqrt{\pi })\sigma }{2\epsilon _{0}n(\nu )cA}}.
\end{equation}
Note that even if group dispersion were included in (\ref{disp}): 
\begin{equation}\label{dispersion}
k(\omega )=k(\nu )+(\omega -\nu )/v_{g}+\frac{1}{2}\kappa _{\nu }(\omega
-\nu )^{2},
\end{equation}%
with $\kappa _{\nu }=(d^{2}k/d\omega ^{2})_{\omega =\nu }$, we would still
obtain the same relation between $E_{0}$ and $U_{em}$. 

\section{The interaction Hamiltonian for a $\protect\chi _{2}$ process}\label{chi2app} 
\noindent We will be using the notation and conventions introduced by Yang et al. \cite{Yang08}. The nonlinear Hamiltonian in the Schr\"{o}dinger picture (equation 13 of \cite{Yang08}) is written as 
\begin{eqnarray}\label{HIintroduce} 
H_{I}&=&-\int_{k}dk_{a}dk_{b}dk_{c}S_{I}(k_{c},k_{a},k_{b})c(k_{c})a^{\dagger
}(k_{a})b^{\dagger }(k_{b})  \nonumber \\
&&+\text{h.c.,}
\end{eqnarray}
where $[c(k),c^{\dagger }(k^{\prime })]=[b(k),b^{\dagger }(k^{\prime
})]=[a(k),a^{\dagger }(k^{\prime })]=\delta (k-k^{\prime })$ and any other
commutator between operators equal to zero. \ We assume that the wavenumber
ranges and/or the polarizations of the pump photons (destruction operator $%
c(k_{c})$) and the down-converted photons (destruction operators $a(k_{a})$
and $b(k_{b})$) are distinct. \ Letting the frequency associated with $k_{l}$
be $\omega _{l}$, where $l=a,b,$ or $c$, the coupling coefficient $%
S_{I}(k_{c},k_{a},k_{b})$ is given by 
\begin{eqnarray}
S_{I}(k_{c},k_{a},k_{b}) &=&\frac{2}{\epsilon _{0}}\sqrt{\frac{\hbar \omega
_{c}\hbar \omega _{a}\hbar \omega _{b}}{(4\pi )^{3}}}\frac{\chi _{2}}{%
\epsilon _{0}n_{c}^{2}\left( \omega _{c}\right) n_{a}^{2}(\omega
_{a})n_{b}^{2}(\omega _{b})} \notag\\
&&\int_{A}dxdy\;d_{c}(x,y)\bar{d}_{a}(x,y)\bar{d}%
_{b}(x,y)\times   \label{Suse} \notag \\
&&\int_{-L/2}^{L/2}dz\;e^{i(k_{c}-k_{a}-k_{b})z},
\end{eqnarray}%
where $n_{l}(\omega _{l})$ is effective index of the indicated mode type. \
Here we have simplified the corresponding expression of Yang et al. \cite%
{Yang08} by neglecting the tensor nature of $\chi _{2}$ and the vector
nature of the $d_{l}(x,y)$, which characterize the mode profiles of the
displacement field; in our expression (\ref{Suse}) for $%
S_{I}(k_{c},k_{a},k_{b})$ the parameter $\chi _{2}$ can be taken to
characterize the strength of the nonlinearity, which we assume to be present
only from $-L/2$ to $L/2,$ and in the spirit of our approximations in the
previous appendix we take the $d_{l}(x,y)$ to be uniform over an area $A$.
\ An extra factor of $2$ appears in (\ref{Suse}) compared with the
corresponding Eq. (13) of Yang et al. \cite{Yang08} because here we assume
that the creation operators $a^{\dagger }(\omega _{a})$ and $b^{\dagger
}(\omega _{b})$ involve different ranges of frequency and/or polarization.
Since the actual vector $\mathbf{d}_{l}(x,y)$ at $\omega _{l}$ are
normalized according to  (equation 8 of\cite{Yang08}) 
\begin{equation*}
\int dxdy\frac{\mathbf{d}_{l}(x,y)\cdot \mathbf{\bar{d}}_{l}(x,y)}{\epsilon
_{0}n_{l}^{2}(\omega _{l})}\frac{v_{p}(l)}{v_{g}(l)}=1,
\end{equation*}%
where $v_{p}(l)=c/n_{l}(\omega _{l})$ and $v_{g}(l)=d\omega _{l}/dk_{l}$ are
the phase and group velocities, this leads to 
\begin{equation}
d_{l}=\sqrt{\frac{v_{g}(l)n_{l}^{2}(\omega _{l})\epsilon _{0}}{v_{p}(l)A}},
\label{dluse}
\end{equation}%
and using this in (\ref{Suse}) we have 
\begin{eqnarray}
S_{I}(k_{c},k_{a},k_{b}) &=&2L\sqrt{\frac{\hbar \omega _{c}\hbar \omega
_{a}\hbar \omega _{b}}{(4\pi )^{3}}}\chi _{2}\\
&&
\sqrt{\frac{%
v_{g}(a)v_{g}(b)v_{g}(c)}{\epsilon
_{0}Av_{p}(a)v_{p}(b)v_{p}(c)n_{a}^{2}n_{b}^{2}n_{c}^{2}}}  \notag \\
&&\times \text{ sinc}\left( \frac{L}{2}(k_{c}-k_{a}-k_{b})\right) \notag .
\end{eqnarray}%
For fields propagating in a single direction it is conventional to work with
frequency rather than wavenumber labels; hence we take, for example, $%
c(k_{c})\rightarrow \sqrt{d\omega _{c}/dk_{c}}c(\omega _{c})$, such that $%
\left[ c(\omega _{a}),c^{\dagger }(\omega _{a}^{\prime })\right] =\delta
(\omega _{a}-\omega _{a}^{\prime })$ (see, e.g., equation 50 in \cite{Yang08}%
). \ Then changing to frequency variables in (\ref{Suse}) we have 
\begin{eqnarray}
H_{I}(t) &=&-2\int_{0}^{\infty }d\omega _{a}d\omega _{b}d\omega _{c} L \chi _{2} c(\omega _{c})a^{\dagger }(\omega
_{a})b^{\dagger }(\omega _{b})e^{i\Delta t} \notag \\
&& \sqrt{\frac{\hbar \omega _{c}\hbar \omega _{a}\hbar \omega _{b}}{(4\pi
)^{3}\epsilon _{0}Ac^{3}n_{a}(\omega _{a})n_{b}(\omega _{b})n_{c}(\omega
_{c})}}  \label{HIoft} \\
&& \text{sinc}\left( \frac{L}{2}(k(\omega _{c})-k(\omega
_{a})-k_{b}(\omega _{b}))\right) \notag +\text{h.c.} ,
\end{eqnarray}%
where we have also moved to the interaction picture defined by linear
evolution, $H_{I}\rightarrow H_{I}(t)$, with $c(\omega _{c})a^{\dagger
}(\omega _{a})b^{\dagger }(\omega _{b})$ picking up the phase factor $\exp
(i\Delta t)$, where $\Delta =\omega _{a}+\omega _{b}-\omega _{c}$. In the
same interaction picture the electric field operator can be related to the
displacement field operator in the usual way; using the expression for the
displacement field operator (Eq. (4) of \cite{Yang08}) and our expression (%
\ref{dluse}) for the mode profile function we find 
\begin{equation*}
\mathbf{E}(\mathbf{r},t)=\sum_{l}\int_{0}^{\infty }\frac{d\omega _{l}}{n_{l}}\sqrt{\frac{\hbar \omega _{l}}{4\pi \epsilon _{0}v_{p}(l)A}}q_{l}(\omega_{l})e^{ik(\omega _{l})z-i\omega _{l}t} +\text{h.c.}
\end{equation*}
where as $l$ is summed over $a,b,$ and $c$, $q_{l}(\omega _{l})$ ranges over
the destruction operators $a(\omega _{a})$, $b(\omega _{b})$, and $c(\omega
_{c}).$ Taking the expectation value of the above operator with respect to a
coherent state for the pump field (and vacuum for the down-converted photon
frequencies) turns $c(\omega _{c})$ into a $c$-number $\beta (\omega )$,
where we drop the subscript $c$. \ Now we can compare this with a classical pulse with energy $U_{em}$, centered at $\nu _{l}$ and frequency variance $\sigma ^{2}$ (\ref{pulseE})
\begin{eqnarray}
\mathbf{E}(\mathbf{r},t) &=&E_{0}\int_{0}^{\infty }d\omega e^{i(k(\omega)z-\omega t)}\frac{1}{\sqrt{2\pi }\sigma }e^{-(\omega -\nu _{l})^{2}/2\sigma^{2}}  \notag \\
&=&\int_{0}^{\infty }d\omega \sqrt{\frac{(U_{em}/\sqrt{\pi })\sigma }{2\epsilon _{0}n(\omega )cA}}e^{i(k(\omega )z-\omega t)}  \notag \\
&&\times \frac{1}{\sqrt{2\pi }\sigma }e^{-(\omega -\nu _{l})^{2}/2\sigma
^{2}},
\end{eqnarray}
and identify
\begin{equation*}
\beta (\omega )=\sqrt{\frac{U_{em}/\sqrt{\pi }}{\hbar \omega \sigma }}e^{-(\omega -\nu )^{2}/2\sigma ^{2}}.
\end{equation*}
It is convenient to replace $\sigma $ by with the inverse standard deviation of the time dependent pump intensity, $\tau =1/\sqrt{2}\sigma =2\sqrt{2\ln 2} \text{ FWHM},$ and write 
\begin{equation}\label{rep}
\beta (\omega )=\sqrt{\frac{U_{em}/\sqrt{\pi /2}}{\hbar \tau \omega }}\tau e^{-\tau ^{2}(\omega -\nu )^{2}}.
\end{equation}
With this in hand we can return to the nonlinear Hamiltonian (\ref{HIoft})
and write 
\begin{eqnarray}
H_{I}(t) &=&-\hbar \varepsilon \int_{0}^{\infty }d\omega _{a}d\omega
_{b}d\omega _{c}\frac{\tau }{\sqrt{\pi }}e^{-\tau ^{2}(\omega _{c}-\nu
_{c})^{2}}\times   \label{HIresult} \\
&&\text{ sinc}\left( \frac{L}{2}(k(\omega _{c})-k_{a}(\omega
_{a})-k_{b}(\omega _{b}))\right) \notag \\
&& e^{i\Delta t}a^{\dagger }(\omega_{a})b^{\dagger }(\omega _{b})+\text{h.c.}  \notag,
\end{eqnarray}%
for an SPDC process with a classically described undepleted pump mode $c$,
where we introduced the dimensionless perturbation parameter $\varepsilon $ 
\begin{equation*}
\varepsilon =2L\chi _{2}\sqrt{\frac{\pi \nu _{b}\nu _{a}}{(4\pi
)^{3}\epsilon _{0}Ac^{3}n_{a}(\nu _{a})n_{b}(\nu _{b})n_{c}(\nu _{c})}}\sqrt{\frac{U_{em}/\sqrt{\pi /2}}{\tau }},
\end{equation*}%
and in benign frequency factors we have evaluated $\omega _{c}$ at $\nu _{c}$
, and $\omega _{a}$ and $\omega _{b}$ at the frequencies $\nu _{a}$ and $\nu
_{b}$ where the phase-matching condition is assumed to be satisfied. 
Defining
\begin{equation*}
\alpha (\omega _{c})=\frac{\tau }{\sqrt{\pi }}e^{-\tau ^{2}(\omega _{c}-\nu
_{c})^{2}},
\end{equation*}
which has units of time, and changing the notation for the pump frequency
from $\omega _{c}$ to $\omega _{p}$, (\ref{HIresult}) reduces to (23), with $%
\Phi (\omega _{a},\omega _{b},\omega _{p})$ given by (26). 

\section{The interaction Hamiltonian for a $\protect\chi _{3}$ process}\label{chi3app} 
\noindent For a third order process, the straightforward extension of
the nonlinear Hamiltonian (\ref{HIintroduce}) \cite{Yang08} in the Schr\"{o}%
dinger picture is given by 
\begin{equation}
H_{I}=-\frac{1}{4\epsilon _{0}}\int_{V}d\mathbf{r\;}\Gamma _{3}^{ijkl}D^{i}(%
\mathbf{r})D^{j}(\mathbf{r})D^{k}(\mathbf{r})D^{l}(\mathbf{r}),
\label{HFWMstart}
\end{equation}%
where indices refer to Cartesian components and are to be summed over when
repeated; $\Gamma _{3}^{ijkl}$ can be written in terms of the more usual
third order optical susceptibility $\chi _{3}^{ijkl}$ and the linear indices
at the appropriate frequency components involved \cite{luke}; as in the appendix
above we use the symbol $\chi _{3}$ (without tensor indices) to indicate the
size of the component(s) that are important when the relevant mode fields
are used in the different $D^{i}(\mathbf{r})$. \ For SFWM we require
destruction operators for the pump fields ($c(\omega _{c})$ and $c(\omega
_{c}^{\prime })$) and creation operators for the photons generated described
by $a\left( \omega _{a}\right) $ and $b(\omega _{b})$, and in analogy with
the result of Appendix B we find 
\begin{eqnarray}\label{HIwork}
H_{I}(t) &=&-\int_{-L/2}^{L/2}dz\int_{0}^{\infty }d\omega _{c}d\omega
_{c}^{\prime }d\omega _{b}d\omega _{a} \\
&& \frac{3\chi _{3}}{\epsilon
_{0}^{3}n_{c}^{2}(\omega _{c})n_{c}^{2}(\omega _{c}^{\prime
})n_{b}^{2}(\omega _{b})n_{a}^{2}(\omega _{a})}   \notag \\
&\times &\sqrt{\frac{\hbar \omega _{c}}{4\pi }\frac{\hbar \omega
_{c}^{\prime }}{4\pi }
\frac{\hbar \omega _{a}}{4\pi }\frac{\hbar \omega _{b}}{4\pi }}A\frac{1}{\sqrt{v_{g}(a)v_{g}(b)v_{g}(c)v_{g}(c^{\prime })}}  \notag
\\
&\times &\sqrt{\frac{v_{g}(c^{\prime })n_{c}^{2}(\omega _{c}^{\prime
})\epsilon _{0}}{v_{p}(c^{\prime })A}\frac{v_{g}(c)n_{c}^{2}(\omega
_{c})\epsilon _{0}}{v_{p}(c)A}
}\notag \\
&\times& \sqrt{\frac{v_{g}(a)n_{a}^{2}(\omega _{a})\epsilon
_{0}}{v_{p}(a)A}\frac{v_{g}(b)n_{b}^{2}(\omega _{b})\epsilon _{0}}{v_{p}(b)A}%
}  \notag \\
&\times &a^{\dagger }(\omega _{a})b^{\dagger }(\omega _{b})c(\omega
_{c})c(\omega _{c}^{\prime })e^{-i(\omega_{c}+\omega _{c}^{\prime }-\omega _{a}-\omega _{b})t}  \notag \\
&\times& e^{i(k_{c}(\omega _{c})+k_{c}(\omega_{c}^{\prime })-k_{a}(\omega _{a})-k_{b}(\omega _{b}))z}+\text{h.c.}  \notag
\end{eqnarray}%
The factor of $3$ in the fraction containing $\chi _{3}$ is $(4!/2!)/4$,
with $(4!/2!)$ the number of permutations involving the ordering of the
creating and annihilation operators; note that this is twice the
corresponding factor appearing in \cite{luke} because here we assume that the
creation operators $a^{\dagger }(\omega _{a})$ and $b^{\dagger }(\omega
_{b}) $ involve different ranges of frequency and/or polarization. \ The
refractive index factors in the denominator, and the appearance of $\epsilon
_{0}^{3}$, arise because of the conversion from $\Gamma ^{ijkl}$ to $\chi
_{3}^{ijkl}$ and the prefactor of $\epsilon _{0}^{-1}$ in (\ref{HFWMstart});
compare with the corresponding terms in (\ref{Suse}). The terms of the form $%
\sqrt{\hbar \omega _{l}/4\pi }$ arise from the expansion of the displacement
field in terms of raising and lowering operators, and again find their
correspondence in terms in (\ref{Suse}). \ The factor of $A$ comes from
integrating over the area in our ``toy model'' for our waveguide and its
modes; the factor $\left( v_{g}(a)v_{g}(b)v_{g}(c)v_{g}(c^{\prime })\right)
^{-1/2}$ arises from the change of variables from wave numbers to
frequencies; and the next terms are the expressions for the mode profiles (%
\ref{dluse}). \ \ Supposing that all the frequency prefactors vary slowly
relative to the harmonic dependence in space and time due to the phase
factors, in those prefactors we set $\omega _{c}$ and $\omega _{c}^{\prime }$
at the center frequency $\nu _{c}$ of the pump, and evaluate $\omega _{a}$
and $\omega _{b}$ at the frequencies $\nu _{a}$ and $\nu _{b}$ where the
phase-matching is optimum; (\ref{HIwork}) then reduces to 
\begin{eqnarray}
H_{I}(t) &=&-\frac{3\chi _{3}\hbar ^{2}}{\epsilon _{0}A(4\pi )^{2}c^{2}}%
\sqrt{\left( \frac{\nu _{c}}{n_{c}(\nu _{c})}\right) ^{2}\frac{\nu _{a}}{%
n_{a}(\nu _{a})}\frac{\nu _{b}}{n_{b}(\nu _{b})}} \notag\\
&\times &\int_{-L/2}^{L/2}dz\int_{0}^{\infty }d\omega _{c}d\omega
_{c}^{\prime }d\omega _{b}d\omega _{a}  \notag \\
&\times &a^{\dagger }(\omega _{a})b^{\dagger }(\omega _{b})c(\omega
_{c})c(\omega _{c}^{\prime })e^{-i(\omega
_{c}+\omega _{c}^{\prime }-\omega _{a}-\omega _{b})t}  \notag \\
&\times& e^{i(k_{c}(\omega _{c})+k_{c}(\omega_{c}^{\prime })-k_{a}(\omega _{a})-k_{b}(\omega _{b}))z}+\text{h.c.}  
\end{eqnarray}%
As in Appendix B we replace the $c$ operators by classical functions (\ref{rep}) and assume that $\sigma \ll \nu _{c},$ which allows us to
extend the range of the $\omega _{c}$ and $\omega _{c}^{\prime }$ integrals to $-\infty $ to $\infty $, and we find 
\begin{eqnarray*}
H_{I}(t) &=&-\frac{3\chi _{3}\hbar ^{2}}{\epsilon _{0}A(4\pi )^{2}c^{2}}%
\sqrt{\left( \frac{\nu _{c}}{n_{c}(\nu _{c})}\right) ^{2}\frac{\nu _{a}}{%
n_{a}(\nu _{a})}\frac{\nu _{b}}{n_{b}(\nu _{b})}} \\
&&\times \frac{U_{em}/\sqrt{\pi /2}}{\hbar \nu _{c}\sigma } \int_{-L/2}^{L/2}dz\int d\omega _{c}d\omega _{c}^{\prime }d\omega
_{b}d\omega _{a} \\
&&\times a^{\dagger }(\omega _{a})b^{\dagger }(\omega _{b})e^{-(\omega
_{c}-\nu _{c})^{2}/2\sigma ^{2}}e^{-(\omega _{c}^{\prime }-\nu
_{c})^{2}/2\sigma ^{2}} \\
&&\times e^{i(k_{c}(\omega _{c})+k_{c}(\omega _{c}^{\prime })-k_{a}(\omega
_{a})-k_{b}(\omega _{b}))z}e^{-i(\omega _{c}+\omega _{c}^{\prime }-\omega
_{a}-\omega _{b})t} \\
&&+\text{h.c.}
\end{eqnarray*}

To simplify this further we look at the integrals over $\omega _{c}$ and $%
\omega _{c}^{\prime }$,
\begin{eqnarray}
I=\int  &&d\omega _{c}d\omega _{c}^{\prime }e^{i(k_{c}(\omega
_{c})+k_{c}(\omega _{c}^{\prime }))z-i(\omega _{c}+\omega _{c}^{\prime })t}
\label{intqq} \\
&&\times e^{-(\omega _{c}-\nu _{c})^{2}/2\sigma ^{2}}e^{-(\omega
_{c}^{\prime }-\nu _{c})^{2}/2\sigma ^{2}}.  \notag
\end{eqnarray}%
Expanding $k_{c}(\omega _{c})$ and $k_{c}(\omega _{c}^{\prime })$ as was
done earlier (\ref{dispersion}), and introducing $\omega _{\pm }=\omega
_{c}\pm \omega _{c}^{\prime }$, we find that (\ref{intqq}) equals 
\begin{eqnarray}
I&=&\int d\omega _{+}e^{ik_{+}(\omega _{+})z}e^{-i\omega _{+}t}e^{-(\omega
_{+}-2\nu _{c})^{2}/4\sigma ^{2}} \\
&&\times \int \frac{d\omega _{-}}{2}e^{-\frac{\omega_{-}^{2}}{4}\left( \frac{1}{\sigma ^{2}}-i\kappa _{c}z\right) }  \notag,
\end{eqnarray}
where 
\begin{equation}
k_{+}(\omega _{+})=2k_{c}(\nu _{c})+(\omega _{+}-2\nu _{c})/v_{c}+\frac{\kappa _{c}}{4}(\omega _{+}-2\nu _{c})^{2}.  \label{kplus}
\end{equation}
The innermost integral is
\begin{equation}
\int \frac{d\omega _{-}}{2}e^{-\frac{\omega _{-}^{2}}{4}\left( \frac{1}{\sigma ^{2}}-i\kappa _{c}z\right) }=\sqrt{\pi }\sigma g(z),
\end{equation}
where 
\begin{equation*}
g(z)\equiv \frac{1}{\sqrt{1-i\kappa _{c}z\sigma ^{2}}}.
\end{equation*}
We can then finally write the Hamiltonian as
\begin{eqnarray}
H_{I}(t) &=&-\hbar \varepsilon \int d\omega _{+}\int d\omega _{b}\int
d\omega _{a}\tilde{\Phi}(\Delta k)  \label{sfwmeps} \\
&\times &e^{i\tilde{\Delta}t}a^{\dagger }(\omega _{a})b^{\dagger }(\omega
_{b})\tilde{\alpha}(\omega _{+})+\text{h.c.}  \notag,
\end{eqnarray}
where 
\begin{equation*}
\tilde{\alpha}(\omega _{+})\equiv \frac{\tilde{\tau}e^{-\tilde{\tau}%
^{2}(\omega _{+}-2\nu _{c})^{2}}}{\sqrt{\pi }},
\end{equation*}%
(with units of time) characterizes the bandwidth of the pump, and we have
introduced $\tilde{\tau}=1/(2\sigma )$ and $\tilde{\Delta}\equiv \omega
_{a}+\omega _{b}-\omega _{+}$, (note the difference between these quantities
and the corresponding $\tilde{\alpha}$, $\tau $, and $\Delta $ of Appendix
B); the dimensionless perturbation parameter 
\begin{equation*}
\varepsilon =\frac{3\chi _{3}}{\epsilon _{0}A(4\pi )^{2}n_{c}c^{2}}\sqrt{%
\frac{\nu _{a}}{n_{a}}\frac{\nu _{b}}{n_{b}}}(\sqrt{2}U_{em})L\sqrt{\pi }%
2\sigma ,
\end{equation*}%
characterizes the strength of the nonlinear interaction, and the modified
phase matching function is
\begin{equation}
\tilde{\Phi}(\Delta k)\equiv \int_{-L/2}^{L/2}\frac{dz}{L}e^{i(k_{a}(\omega
_{a})+k_{b}(\omega _{b})-k_{+}(\omega _{+}))z}g(z).  \label{phitilde}
\end{equation}%
If the effect of group velocity dispersion on the pump pulse is negligible
over the length $L$ of the nonlinear region, for all relevant $z$ we have $%
\left\vert \kappa _{c}z\sigma ^{2}\right\vert \ll 1$, we can set $%
g(z)\rightarrow 1$ and $k_{+}(\omega _{+})=2k_{c}(\omega _{+}/2)$ (cf. (\ref%
{kplus})) and (\ref{phitilde}) reduces to (\ref{pdc}). Finally, note that we can rewrite expression (\ref{sfwmeps}) in a
slightly more suggestive manner, 
\begin{eqnarray}
H_{I}(t) &=&-\hbar \varepsilon \int d\omega _{+}\int d\omega _{b}\int
d\omega _{a}\tilde{\Phi}(\Delta k)  \label{final} \\
&\times &e^{i\tilde{\Delta}t}a^{\dagger }(\omega _{a})b^{\dagger }(\omega
_{b})\frac{\zeta (\omega _{+}/2-\nu _{c})^{2}}{2\sigma }+\text{h.c.}  \notag
\end{eqnarray}%
where $\zeta (\omega )$ is proportional to the electric field amplitude as
defined in (\ref{disp}), explicitly displaying the dependence of $H_{I}(t)$
on the square of that amplitude. \ Of course, the details of the argument
given here rely on the assumed Gaussian nature of the pump.  But we can
expect that the general form (\ref{sfwmeps}) will survive even for more
general pump pulses, as long as the effect of group velocity dispersion on
the pump pulse is negligible. 
\bibliography{main}

\end{document}